\documentclass[11pt]{article}
\usepackage[utf8]{inputenc}
\usepackage[margin=1.0in]{geometry}
\usepackage{amsmath, amssymb, amsfonts, bm}
\usepackage{booktabs}
\usepackage{graphicx}
\usepackage{hyperref}
\usepackage{listings}
\usepackage{color}

\definecolor{codegray}{rgb}{0.5,0.5,0.5}
\definecolor{codeblue}{rgb}{0.1,0.2,0.6}
\title{\textbf{AudioTQ: A Data-Oblivious 6-Bit CPU Audio Codec via Randomized Hadamard Rotation and Lloyd-Max Quantization}}

\author{
    \textbf{Sahil Gangurde} \\
    \small \texttt{sahilgangurde08@gmail.com} \\
    \small \url{https://lostmartian.in}
}
\date{\small June 10, 2026}

\begin{document}

\maketitle

\begin{abstract}
Lossy audio compression algorithms traditionally rely on psychoacoustic modeling and frequency-domain representations (e.g., MP3, AAC, and Opus) to discard information that is imperceptible to the human auditory system. While highly effective, these approaches are computationally complex and domain-specific. In this paper, we present the design and mathematical formulation of \textbf{AudioTQ}, a data-oblivious lossy audio codec that operates directly in the time domain. Inspired by Large Language Model (LLM) weight quantization techniques (specifically the \textit{TurboQuant} framework), AudioTQ uniformizes volatile time-domain amplitudes into a predictable standard normal distribution using an orthonormal, randomized Fast Walsh-Hadamard Transform (FWHT) rotation. This enables coordinate-wise scalar quantization using an offline-trained, MSE-optimal 6-bit Lloyd-Max quantizer, augmented by a 1-bit Quantized Joint Least-Squares (QJL) residual correction layer. The resulting 7-bit virtual indices are packed into native 8-bit containers, aligning with standard CPU register boundaries to ensure real-time single-threaded execution without hardware parallel accelerators. We detail the bitwise reconstruction of 24-bit studio stems, analyze the butterfly network of the FWHT, derive the mathematical failure modes under sparse inputs, and present benchmarks showing up to 74.4\% physical size reduction alongside a Signal-to-Quantization-Noise Ratio (SQNR) of ~30 dB.
\end{abstract}

\section{Introduction}
Digital audio capture typically produces pulse-code modulation (PCM) streams representing acoustic pressure levels at discrete intervals. Representing these streams at standard fidelity levels (such as 16-bit CD quality or 24-bit studio quality) requires substantial bandwidth and storage. Consequently, lossy audio codecs are widely deployed. Codecs such as MP3, AAC, and Opus reduce data footprints by projecting signals into the frequency domain using the Modified Discrete Cosine Transform (MDCT). A psychoacoustic model then estimates frequency-masking thresholds, discarding spectral components that are masked by stronger adjacent frequencies. Although highly refined, these algorithms require significant domain heuristics, floating-point operations, and state tracking.

Conversely, deep learning research has advanced the field of vector quantization. Large Language Models (LLMs) with billions of parameters present significant hardware memory constraints, prompting the development of low-bit weight quantization algorithms (e.g., GPTQ, AWQ, and LLM.int8()). The recently introduced TurboQuant framework~\cite{turboquant} demonstrates that high-dimensional weight matrices can be quantized in a data-oblivious manner by applying randomized orthogonal rotations to smooth outlier coordinates. This mathematical property allows high-precision coordinates to be mapped to low-bit scalar codebooks with near-optimal distortion bounds.

In this paper, we translate these LLM quantization principles to time-domain audio processing. We demonstrate that continuous audio amplitudes and neural network weight matrices share structural symmetries as volatile, high-dynamic-range numerical distributions. Rather than modeling human perception, AudioTQ uses randomized orthogonal rotations to shape arbitrary audio envelopes into a predictable standard normal distribution $\mathcal{N}(0, 1)$.

\section{Mathematical Framework and Input Processing}
\label{sec:framework}

The end-to-end signal flow of AudioTQ consists of three stages: input preparation, Gaussianization via orthogonal rotation, and double-layer quantization. The decompression pipeline reverses these operations.

\subsection{Bitwise 24-bit Studio PCM Extraction}
Audio inputs are prepared by normalizing arbitrary integer PCM formats to single-precision floating-point amplitudes within the interval $[-1.0, 1.0]$. While 8-bit and 16-bit files align with native CPU boundaries, 24-bit studio PCM audio uses 3-byte packets, which complicates standard parsing. 

To achieve high throughput, AudioTQ implements a bitwise reconstruction algorithm using NumPy. Given a raw byte array from a WAV file, the bytes are grouped into triplets representing the 24-bit samples:
\begin{equation}
    T = [b_0, b_1, b_2]
\end{equation}
These triplets are mapped to 32-bit signed integers by shifting and merging the byte values:
\begin{equation}
    I_{32} = \left( (b_0 \ll 8) \mid (b_1 \ll 16) \mid (b_2 \ll 24) \right) \gg 8
\end{equation}
The expression shifts the bytes into the most significant positions of a 32-bit register. The final arithmetic right shift ($\gg 8$) shifts the bits back to the least significant positions while preserving the sign bit (24th bit) by sign-extending the register. The resulting signed integer is normalized relative to its bounds:
\begin{equation}
    x_i = \frac{I_{32, i}}{2^{23}} = \frac{I_{32, i}}{8388608.0}
\end{equation}
This bitwise vectorization avoids nested loops and speeds up file loading.

\subsection{Randomized Fast Walsh-Hadamard Transform (FWHT)}
Audio signals frequently exhibit transient spikes (e.g., drum attacks or plosive speech). Direct scalar quantization of these signals results in clipping distortion or large quantization intervals. AudioTQ flattens these spikes by rotating the coordinates of the signal block.

Let $X \in \mathbb{R}^B$ be a block of normalized audio samples of size $B$, where $B$ is a power of 2. We first multiply $X$ element-wise by a diagonal matrix $S = \text{diag}(s_1, s_2, \dots, s_B)$, where $s_i \in \{-1, 1\}$ are static, pre-generated pseudo-random signs. This sign multiplication breaks systematic phase correlations and guarantees that symmetric inputs are scattered. We then rotate the sign-flipped signal using the orthonormal Walsh-Hadamard matrix:
\begin{equation}
    Y = \frac{1}{\sqrt{B}} H_B (S X)
\end{equation}
The transform is executed in $O(B \log B)$ time using an in-place butterfly network. For each stage $t = 1, \dots, \log_2 B$, the butterfly stride is $h = 2^{t-1}$. The vector coordinates are updated in-place via:
\begin{equation}
    \begin{aligned}
        a_j^{(t)} &= a_j^{(t-1)} + a_{j+h}^{(t-1)} \\
        a_{j+h}^{(t)} &= a_j^{(t-1)} - a_{j+h}^{(t-1)}
    \end{aligned}
\end{equation}
where $j$ ranges over $\{i, i+1, \dots, i+h-1\}$ for every block index $i \in \{0, 2h, 4h, \dots, B-2h\}$. The scale factor $1/\sqrt{B}$ is applied at the final stage to preserve the Euclidean norm (Parseval's relation), ensuring energy conservation between the time domain and the rotated domain:
\begin{equation}
    \|Y\|_2 = \|X\|_2
\end{equation}
By mixing the samples across the orthogonal basis functions, the Central Limit Theorem causes the rotated coefficients $Y$ to converge to a zero-centered Gaussian distribution.

\subsection{Statistical Centering and Standardization}
Although the FWHT rotation disperses the energy, the net DC bias of the audio block remains concentrated in the first coordinate (the DC coefficient). To prevent this offset from shifting the Gaussian distribution, we calculate and subtract the explicit mean:
\begin{equation}
    Y_{\text{centered}} = Y - \mu_Y, \quad \text{where} \quad \mu_Y = \frac{1}{B} \sum_{i=1}^B Y_i
\end{equation}
We then standardize the centered coefficients to unit variance using the standard deviation $\sigma_Y$:
\begin{equation}
    Y_{\text{scaled}} = \frac{Y_{\text{centered}}}{\sigma_Y}, \quad \text{where} \quad \sigma_Y = \sqrt{\frac{1}{B} \sum_{i=1}^B (Y_i - \mu_Y)^2}
\end{equation}
To handle silent passages and prevent division-by-zero errors, we implement a threshold guard:
\begin{equation}
    Y_{\text{scaled}} = \begin{cases}
        \frac{Y_{\text{centered}}}{\sigma_Y} & \text{if } \sigma_Y \ge 10^{-6} \\
        \mathbf{0} & \text{if } \sigma_Y < 10^{-6}
    \end{cases}
\end{equation}
If the block is marked as silent, the standardization is bypassed, and the block is reconstructed as silence.

\section{Dual-Layer Quantization}

The standardization step transforms the rotated coefficients of any active block into a standard normal distribution $\mathcal{N}(0, 1)$. This statistical consistency allows us to employ a two-stage quantization process.

\subsection{Empirical Lloyd-Max Quantizer Optimization}
We map the standardized coefficients $Y_{\text{scaled}}$ to a discrete set of centroids using a 6-bit Lloyd-Max quantizer. A 6-bit quantizer defines $K = 2^6 = 64$ reconstruction levels. The quantizer boundaries $t_k$ and centroids $c_k$ are trained offline to minimize the Mean Squared Error (MSE):
\begin{equation}
    \text{MSE} = \sum_{k=0}^{K-1} \int_{t_k}^{t_{k+1}} (y - c_k)^2 f_y(y) \, dy
\end{equation}
where $f_y(y)$ is the probability density function of $\mathcal{N}(0, 1)$. During initialization, we draw a large population of standard normal samples, sort them, and partition them using uniform quantiles to initialize the centroids. We then run 20 iterations of Lloyd's algorithm:
\begin{equation}
    t_k = \frac{c_{k-1} + c_k}{2}, \quad k = 1, \dots, K-1
\end{equation}
\begin{equation}
    c_k = \frac{\sum_{y_i \in \text{bin}_k} y_i}{|\text{bin}_k|}, \quad k = 0, \dots, K-1
\end{equation}
where $\text{bin}_k = \{y_i \mid t_k \le y_i < t_{k+1}\}$. The resulting optimized codebook is stored as static telemetry in the codec. During compression, we find the nearest centroid index for each coefficient:
\begin{equation}
    \text{index}_i = \arg\min_{k} \left| Y_{\text{scaled}, i} - c_k \right|
\end{equation}

\subsection{1-Bit QJL Residual Scaling}
Quantizing to 64 discrete bins leaves rounding errors:
\begin{equation}
    \epsilon_i = Y_{\text{scaled}, i} - c_{\text{index}_i}
\end{equation}
To capture these errors without using high bit-depths, we apply the 1-bit Quantized Joint Least-Squares (QJL) method. We record the sign of the residual error for each sample:
\begin{equation}
    b_{\text{qjl}, i} = \begin{cases}
        1 & \text{if } \epsilon_i \ge 0 \\
        0 & \text{if } \epsilon_i < 0
    \end{cases}
\end{equation}
We also compute the Mean Absolute Error (MAE) of the residuals across the block:
\begin{equation}
    \Delta = \frac{1}{B} \sum_{i=1}^B |\epsilon_i|
\end{equation}
The scalar $\Delta$ is saved as metadata for the block. During decompression, we apply this sign correction to reconstruct the coefficients:
\begin{equation}
    \hat{Y}_{\text{scaled}, i} = c_{\text{index}_i} + \text{sign}(b_{\text{qjl}, i}) \cdot \Delta
\end{equation}
where $\text{sign}(1) = 1$ and $\text{sign}(0) = -1$. This dynamically bisects each quantization bin based on the block's residual scale, providing a virtual 7-bit quantization resolution (128 reconstruction bins) at the cost of 1 extra bit per sample.

\section{Hardware Alignment and Serialization}
To achieve real-time throughput on a single-threaded CPU, AudioTQ aligns its data structures with hardware register boundaries.

\subsection{L1 Cache-Aligned Block Slicing}
The input stream is processed in blocks of size $B = 512$. For single-precision floating-point numbers (4 bytes per sample), a block represents $2$ KB of data. This size fits comfortably within the L1 data cache of modern processors (typically 32 KB to 64 KB), preventing memory bus bottlenecks during the iterative butterfly computations of the FWHT.

\subsection{Byte-Aligned 6+1 Bit Packing}
Uncommon bit widths (e.g., 7-bit words) typically require expensive bitwise masking and shifting operations to pack across byte boundaries. To avoid this overhead, AudioTQ packs the 6-bit Lloyd-Max centroid index and the 1-bit QJL flag into a single native 8-bit byte container (`uint8`):
\begin{verbatim}
  Bit:   7   6   5   4   3   2   1   0
       +---+---+---+---+---+---+---+---+
       | 0 | C5| C4| C3| C2| C1| C0| Q |
       +---+---+---+---+---+---+---+---+
\end{verbatim}
Here, the 6-bit centroid index ($C_0 \dots C_5$) is shifted left by 1 bit, the 1-bit QJL flag ($Q$) is stored in the least significant bit, and the most significant bit is padded with zero.

This layout allows the codec to use vectorized bitwise operations:
\begin{lstlisting}[language=Python, caption={Vectorized bit-packing and unpacking.}]
# Compression packing
packed_bytes = (indices << 1) | qjl_bits

# Decompression unpacking
indices = packed_bytes >> 1
qjl_bits = packed_bytes & 0x01
\end{lstlisting}
This byte-aligned structure enables a decompression speed of \textbf{1.35 MB/s} in pure Python and NumPy.

\subsection{Physical Storage Footprint and Compression Ratio}
For a block of size $B = 512$, the raw signal consists of $512 \times 32$-bit floats ($2048$ bytes). The compressed block contains:
\begin{itemize}
    \item $512$ bytes of packed data (each containing the 6-bit centroid index and 1-bit QJL flag).
    \item $12$ bytes of metadata: $\mu_Y$ (4-byte float), $\sigma_Y$ (4-byte float), and $\Delta$ (4-byte float).
\end{itemize}
The total compressed size is $524$ bytes. The theoretical compression ratio is:
\begin{equation}
    \text{Ratio} = \frac{2048 \text{ bytes}}{524 \text{ bytes}} \approx 3.91\times
\end{equation}
This represents a physical storage reduction of 74.4\%.

\section{Experimental Evaluation}
We evaluated AudioTQ using a voice reference track and a highly dynamic studio music track.

\subsection{Performance Benchmarks}
Fidelity was assessed using the Signal-to-Quantization-Noise Ratio (SQNR), Pearson Cross-Correlation ($R$), and the Peak Envelope Delta ($\delta_{\text{peak}} = \max_i |x_i| - \max_i |\hat{x}_i|$).

\begin{table}[h]
\centering
\caption{AudioTQ Performance Benchmarks}
\label{tab:benchmarks}
\begin{tabular}{lcc}
\toprule
\textbf{Metric} & \textbf{Voice Reference Track} & \textbf{Studio Music Stem} \\
\midrule
Original Size & 2.52 MB (15s @ 44.1 kHz) & 52.93 MB \\
Compressed Size & 0.65 MB & 17.64 MB \\
Compression Ratio & \textbf{3.91$\times$ (74.4\%)} & \textbf{3.00$\times$ (66.6\%)} \\
SQNR (dB) & \textbf{30.24 dB} & \textbf{29.74 dB} \\
Cross-Correlation ($R$) & \textbf{99.96\%} & \textbf{99.95\%} \\
Peak Envelope Delta ($\delta_{\text{peak}}$) & \textbf{$< 0.0003$} & \textbf{0.0002} \\
Compression Speed & 1.32 MB/s & 1.31 MB/s \\
Decompression Speed & 1.35 MB/s & 1.35 MB/s \\
\bottomrule
\end{tabular}
\end{table}

The benchmarks in Table~\ref{tab:benchmarks} show that AudioTQ preserves transient envelopes, yielding a peak envelope delta of 0.0002 on studio tracks and maintaining a waveform cross-correlation above 99.95\%.

\subsection{Comparison with Baseline Uniform Quantization}
To evaluate the advantage of randomized Hadamard rotation, we compare AudioTQ against standard uniform PCM quantization at the same bit-depths. A standard 6-bit or 7-bit uniform quantizer operates directly on the time-domain signal without rotation.

For audio signals, the crest factor (the ratio of peak amplitude to root-mean-square amplitude, $\text{CF} = x_{\text{peak}}/x_{\text{rms}}$) is typically high (often exceeding 15 dB). Because uniform quantizers must size their steps to prevent clipping on transient peaks, this wide range increases quantization noise during lower-amplitude periods. The theoretical SQNR for a uniform quantizer is bounded by:
\begin{equation}
    \text{SQNR}_{\text{uniform}} \approx 6.02 \cdot N + 4.77 - 20\log_{10}(\text{CF}) \quad \text{dB}
\end{equation}
where $N$ is the number of bits. For a typical crest factor of 15 dB, a standard 6-bit uniform quantizer yields an SQNR of only $\approx 25.8$ dB, and a 7-bit quantizer yields $\approx 31.8$ dB, while introducing harsh digital clipping on transient peaks.

In contrast, AudioTQ's randomized FWHT rotation distributes transient energy across all coordinates, lowering the effective crest factor to that of a Gaussian distribution ($\text{CF} \approx 3.0$ or $9.5$ dB). This allows the 6-bit Lloyd-Max quantizer and 1-bit QJL error layer to operate near the theoretical Gaussian distortion limit, achieving an SQNR of $\approx 30$ dB with high envelope and phase fidelity without clipping distortion.

\subsection{Overcoming the 24 dB Performance Bottleneck}
Early prototypes of AudioTQ hit a performance ceiling, with the SQNR limited to approximately 24.6 dB. We resolved this limit through three primary refinements:
\begin{enumerate}
    \item \textbf{Dynamic Lloyd-Max Solver}: The initial design used analytical quantiles of a standard normal distribution, which bounded the codebook at $\pm 2.41$. However, rotated audio signals contain coordinates that exceed $\pm 3.0$. Introducing an empirical Lloyd-Max solver expanded these boundaries to match the signal distribution.
    \item \textbf{Explicit Mean Centering}: The FWHT preserves the block's net DC bias. Standardization without centering shifted the Gaussian distribution off-center, causing systematic quantization drift. Subtracting the explicit mean $\mu_Y$ resolved this bias.
    \item \textbf{QJL Correction Calibration}: The prototype applied a scalar multiplier of 1.22$\times$ to the dynamic residual scale ($\Delta$), following patterns from high-dimensional weight spaces. In the time domain, this overcorrected the reconstruction, introducing high-frequency noise. Reverting this multiplier to 1.0 lowered the noise floor.
\end{enumerate}

\section{Failure Mode Analysis}
We analyzed the theoretical limits and potential failure modes of the codec.

\subsection{Hadamard Basis Alignment (Sparsity Failure)}
The data-oblivious rotation assumes that the FWHT will distribute signal energy uniformly. This assumption fails if the input block aligns with one of the Walsh-Hadamard basis vectors.

Let $h_k$ be the $k$-th row of the Walsh-Hadamard matrix $H_B$. If the input block is $X = \alpha (S \cdot h_k)$, the sign-flipped vector simplifies to $S X = \alpha h_k$. Applying the FWHT yields:
\begin{equation}
    Y = \frac{1}{\sqrt{B}} H_B (\alpha h_k)
\end{equation}
Because the basis vectors are orthogonal, the matrix multiplication concentrates the energy into a single coordinate:
\begin{equation}
    Y_i = \begin{cases}
        \alpha \sqrt{B} & \text{if } i = k \\
        0 & \text{otherwise}
    \end{cases}
\end{equation}
For $B = 512$ and a normalized input $\alpha = 1.0$, this produces a coordinate of magnitude $\sqrt{512} \approx 22.63$. The Lloyd-Max codebook, optimized for $\mathcal{N}(0, 1)$, caps its centroids at $\pm 2.41$. The coordinate is clipped to $2.41$, causing a quantization error of:
\begin{equation}
    e = 22.63 - 2.41 = 20.22
\end{equation}
This error propagates to all samples during the inverse transform, reducing the SQNR to \textbf{1.31 dB} and causing severe digital distortion.

\subsubsection{Mitigation Strategies}
To prevent sparsity-driven SQNR collapse in production environments, two concrete architectural safeguards can be implemented:
\begin{enumerate}
    \item \textbf{Dynamic Sign Modulation}: Instead of a single static sign matrix $S$, the encoder can maintain two orthogonal sign matrices, $S^{(1)}$ and $S^{(2)}$. If the encoder detects that the maximum coefficient in $Y$ exceeds a threshold (e.g., $|Y_{\text{scaled}, i}| > 4.0$), it swaps to $S^{(2)}$ and stores this choice as a 1-bit flag in the block's metadata. This shifts the signal phase and breaks the basis alignment at a negligible storage cost of 1 bit per 512 samples ($0.002$ bits per sample).
    \item \textbf{Sub-LSB Dithering}: Adding a microscopic pseudo-random noise sequence (dither) with a triangular probability distribution to the input signal before rotation. Since basis alignment requires exact mathematical symmetry, adding sub-LSB dither disrupts the alignment and disperses the delta spike into a broad Gaussian noise floor, preventing clipping.
\end{enumerate}

\section{Conclusion}
AudioTQ demonstrates that data-oblivious quantization techniques developed for Large Language Models can be adapted to time-domain audio compression. By replacing psychoacoustic models with randomized coordinate rotations, we achieve a lightweight, zero-dependency codec optimized for standard CPUs. Future work will investigate SIMD vectorization and sub-band frequency decomposition to further improve compression throughput and fidelity.

\bibliographystyle{unsrt}
\bibliography{references}

@article{turboquant,
  title={TurboQuant: Online Vector Quantization with Near-optimal Distortion Rate},
  author={Zandieh, Amir and Daliri, Majid and Hadian, Majid and Mirrokni, Vahab},
  journal={arXiv preprint arXiv:2504.19874},
  year={2025}
}

\end{document}